\documentclass[aps,prb,preprint,floatfix,superscriptaddress,showpacs]{revtex4-1}

\usepackage{graphicx}
\usepackage{dcolumn}
\usepackage{amsmath,bm}
\usepackage{color}%
\usepackage{multirow}
\usepackage{array} 
\usepackage{xr}
\usepackage{xr-hyper}
\usepackage{hyperref}
\usepackage{chemformula} 
\usepackage[T1]{fontenc} 
\usepackage{braket}
\usepackage{mathtools}
\usepackage{esvect}
\usepackage{amssymb}
\usepackage{tabularx}
\usepackage{soul}
\setstcolor{red}

\newcommand*{\rom}[1]{\expandafter\@slowromancap\romannumeral #1@}
\makeatletter

\newcommand*{\addFileDependency}[1]{
  \typeout{(#1)}
  \@addtofilelist{#1}
  \IfFileExists{#1}{}{\typeout{No file #1.}}
}
\makeatother

\newcommand*{\myexternaldocument}[1]{%
    \externaldocument{#1}%
    \addFileDependency{#1.tex}%
    \addFileDependency{#1.aux}%
}

\myexternaldocument{arxivSI}

\begin{document}

\title{Interfacial Magnetic Anisotropy of Iron-Adsorbed Ferroelectric Perovskites: First-Principles and Machine Learning Study}

\author{Dameul Jeong}
\affiliation{Department of Physics and Research Institute for Basic Sciences, Kyung Hee University, Seoul, 02447, Korea}
\author{Seoung-Hun Kang}
\email[Corresponding author. email: ]{physicsksh@khu.ac.kr}
\affiliation{Department of Physics and Research Institute for Basic Sciences, Kyung Hee University, Seoul, 02447, Korea}
\affiliation{Department of Information Display, Kyung Hee University, Seoul, 02447, Korea}
\affiliation{Materials Science and Technology Division, Oak Ridge National Laboratory, Oak Ridge, TN, 37831, United States of America}
\author{Young-Kyun Kwon}
\email[Corresponding author. email: ]{ykkwon@khu.ac.kr}
\affiliation{Department of Physics and Research Institute for Basic Sciences, Kyung Hee University, Seoul, 02447, Korea}
\affiliation{Department of Information Display, Kyung Hee University, Seoul, 02447, Korea}
\date{May 2025}

\begin{abstract}
The advancement of spin-based devices as a replacement for CMOS technology demands lower spin-switching energy in ferromagnetic (FM) materials. Ferroelectric (FE) materials offer a promising avenue for influencing FM properties, yet the mechanisms driving this interplay remain inadequately understood. In this study, we investigate iron-adsorbed FE ABO$_3$ perovskites using a combination of first-principles calculations and machine learning. Our findings reveal a universal correlation between the magnetic anisotropy energy (MAE) of iron and the induced magnetic dipole moments within the BO$_2$ layer and basal oxygen atoms of ABO$_3$ at the FE/FM interface. By identifying key material descriptors and achieving high predictive accuracy, this research provides a robust framework for selecting and optimizing ABO$_3$ substrates for energy-efficient spintronic devices. These insights contribute to the rational design of novel low-power spin-based technologies.
\end{abstract}

\maketitle

\section{Introduction}
The miniaturization of complementary metal-oxide-semiconductor (CMOS) technology has driven the development of compact and fast electronic devices for decades. However, as device sizes approach 5~nm, quantum tunneling effects hinder the smooth flow of electrons, presenting a critical bottleneck.\cite{wang2006quantum} While extensive efforts have been made to overcome this limitation, a fundamental solution remains elusive.\cite{ratnesh2021advancement,samal2020journey,prasad2019review} Spintronics, an emerging alternative,\cite{wolf2001,vzutic2004spintronics,Fert2008} harnesses the electron’s spin rather than its charge to operate devices, offering revolutionary potential for energy-efficient electronics.
One prominent application of spintronics is spin-transfer torque magnetoresistive random-access memory (STT-MRAM), which delivers near-zero standby power and eliminates leakage currents.\cite{wolf2010,salehi2017survey} Despite these advantages, STT-MRAM requires substantial energy for writing and maintaining magnetic states.\cite{nozaki2019recent} Among various approaches\cite{gupta2019write,eswar2021study,cai2021toward} addressing this challenge, Intel's magnetoelectric spin-orbit (MESO) devices demonstrate significant energy savings, outperforming traditional CMOS devices by a factor of 10 to 30. MESO devices rely on spin-to-charge conversion and magnetization switching via the magnetoelectric effect of multiferroic materials, such as La-doped BiFeOi$_3$.\cite{manipatruni2019scalable} However, the practical adoption of such materials has been limited due to high costs, reliability concerns, and the scarcity of multiferroic compounds exhibiting robust ferroelectric and ferromagnetic properties at room temperature.\cite{yang2015bifeo3,mundy2016atomically}

A promising strategy to overcome these challenges involves reducing the energy required for spin switching by manipulating the magnetic anisotropy energy (MAE) of ferromagnetic (FM) materials. Ferroelectric (FE) materials, with their spontaneous polarization, offer the potential to influence MAE, providing a pathway for lowering spin-switching energy. Previous studies have demonstrated the impact of FE substrates, such as BaTiO$_3$ (BTO) and HfO$_2$, on the MAE of FM materials.\cite{duan2006predicted,Duan2008,Lukashev2012,radaelli2014electric,Odkhuu2017,vermeulen2019ferroelectric,Cheng2023} However, the interplay between the FE substrate properties and FM layer behaviors, particularly at the atomic interface, remains insufficiently understood.

Here, we present our first principles and machine learning investigation of the MAE of iron adsorbates on various FE ABO$_3$ perovskite substrates. These substrates consist of alkaline earth metals (A$=$Ca, Sr, Ba) and group 4 transition metals (B$=$Ti, Zr, Hf). Our study identifies unique factors influencing the MAE of iron adsorbates, providing valuable insights for selecting and optimizing FE and FM materials for spintronic applications. While the MAE of FM layers on FE substrates is well-documented in terms of the spontaneous polarization of FE substrates,\cite{duan2006predicted,Duan2008} our findings reveal that the MAE of iron adsorbates correlates not only with the spontaneous polarization of ABO$_3$ but also with the induced magnetic dipole moments in the BO$_2$ layer and basal oxygen of the ABO$_3$ octahedron near the interface. Employing machine learning, we further explore the universal behavior of FM MAE in the presence of FE layers, demonstrating that the induced magnetic moment near the interface is a critical feature explaining the FE-dependent behavior of FM MAE.

\section{Computational methodology}
We carried out \textit{ab initio} calculations based on density functional theory (DFT),\cite{{hohenberg1964inhomogeneous,kohn1965self}} as implemented in the Vienna \textit{ab initio} simulation package (VASP).\cite{kresse1993ab,kresse1996efficient} Projector augmented wave potentials\cite{blochl1994projector} were employed to ensure the accurate and efficient description of the valence electrons. For the exchange-correlation functional, we used the Ceperley-Alder form\cite{ceperley1980ground} within the local density approximation (LDA). All calculations were performed with an energy cutoff of 500~eV to ensure computational efficiency without sacrificing accuracy.

To obtain reliable structural configurations, we relaxed all structures until the Hellmann-Feynman force on each atom was below 0.01~eV/{\AA}. The Brillouin zone was sampled using $10\times10\times10$ $k$-point grids for bulk structures and $12\times12\times1$ $\Gamma$-centered $k$-point grids for slab structures. We considered the spin-orbit interaction and the Berry phase approach\cite{Spaldin2012} to evaluate the MAE and the spontaneous polarization. 

To examine the universal behaviors of the MAE of iron adsorbates on different ABO$_3$ substrates, we applied the Sure Independence Screening and Sparsifying Operator (SISSO) method.\cite{ouyang2018} SISSO generates an extensive feature space encompassing all measurable quantities related to our primary interest. It then employs the Sure Independence Screening (SIS) technique to select subspaces from their feature space and utilizes the Sparsifying Operator (SO) to achieve sparsity, ultimately providing an optimal $n$-dimensional descriptor. 

During this process, we analyzed a wide range of characteristics representing the essential properties of our Fe-adsorbed ABO$_3$ slab model. We employed both simple mathematical operations (addition, subtraction, multiplication, and division) and more complex functions (exponentials and trigonometric functions) to systematically and recursively create meaningful combinations. This thorough approach resulted in four distinct features with a complexity level of three mathematical operations. From a large collection, we selected a subspace size of 10,000, leading to the discovery of an optimized two-dimensional descriptor that effectively captures the universal behavior of the MAE of iron adsorbed on various ABO$_3$ compounds with full coverage.

It is noted that we employed LDA in all first-principles calculations, primarily due to its numerical stability when combined with spin–orbit coupling and the magnetic force theorem. To validate this choice, we computed the magnetic moment of bulk Fe using LDA, GGA, GGA$+U$ ($U=1-3$~eV), and a hybrid functional. The LDA value (2.141~$\mu_B$) closely matches the experimental value (2.13~$\mu_B$), while GGA and hybrid functionals significantly overestimate it, as summarized in Supplementary Table S1. We further confirmed the robustness of our findings by recalculating MAE using GGA$+U$ for multiple $U$ values. Although the absolute MAE values vary slightly, the main trends, such as the appearance or absence of discrete MAE jumps, remain unchanged. These validations are detailed in Supplementary Note S3 and Supplementary Fig.S4.

\section{Results and discussion}
\begin{figure*}[t]
\centering
\includegraphics[width=0.95\textwidth]{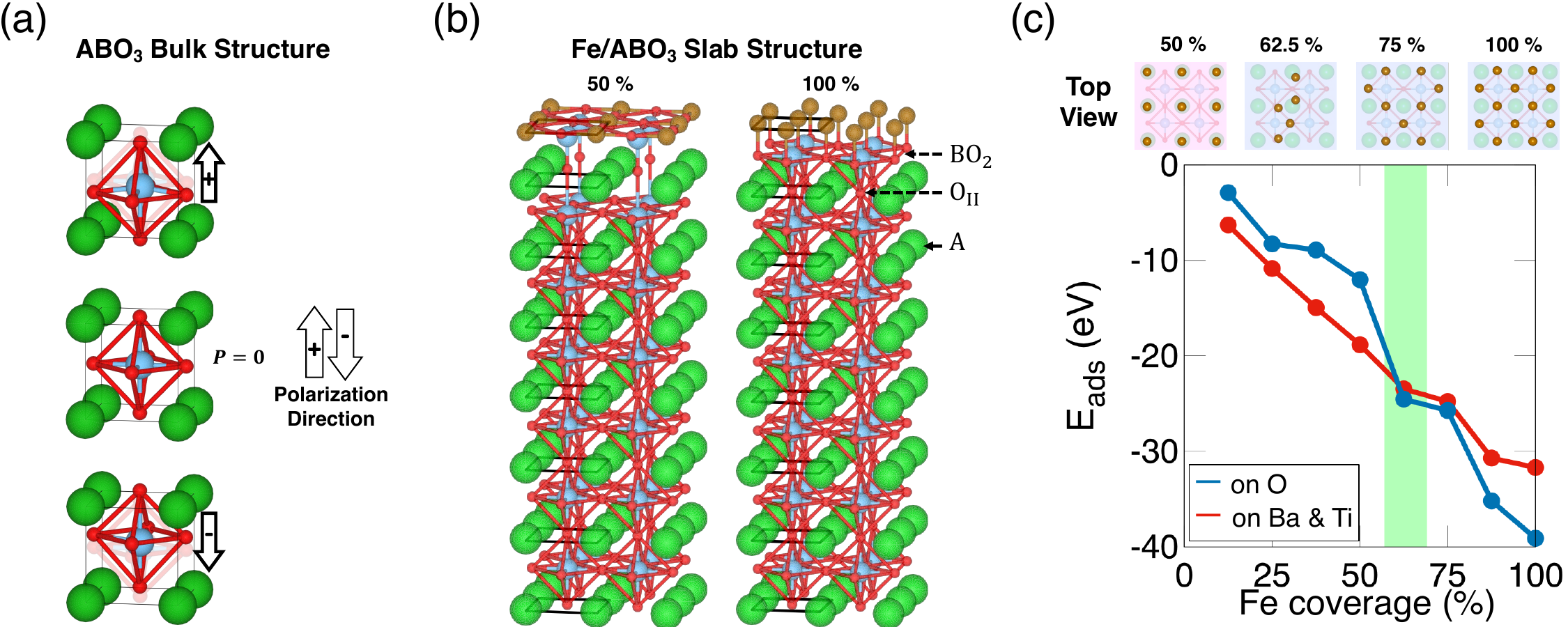}
\caption{(a) Illustrations of three ABO$_3$ bulk perovskite structures displaying different polarization states. The equilibrium structure of bulk ABO$_3$ is tetragonal, inducing spontaneous polarization, shown in the top and bottom configurations with opposite polarization directions. The middle configuration represents the centrosymmetric structure without spontaneous polarization. Green, cyan, and red spheres represent A, B, and O atoms, respectively. (b) Slab models of Fe-adsorbed seven-layer ABO$_3$ substrates with half (50\%) and full (100\%) coverage, respectively. Black squares on the A-atom layer indicate the in-plane unit cell size. The bottom four layers of the ABO$_3$ substrate were fixed to describe bulk effects, while the top three layers were relaxed to describe surface effects. At the interface, there are two nonequivalent oxygen atoms, one is O in the BO$_2$ layer and O$_\mathrm{II}$ just below the topmost B atom.
(c) Adsorption energy $E_\mathrm{ads}$ of iron atoms on the BaTiO$_3$ (BTO) substrate as a function of coverage, where the red and blue curves represent two cases of Fe adsorption on the Ba and O top sites, respectively. The green shaded region indicates the shift of the preferential adsorption sites from the Ba to the O top sites. The top view of stable Fe-adsorbed BTO configurations with four different coverages is also shown above. 
\label{struc}}
\end{figure*}

Perovskite is a mineral with the same crystal form as CaTiO$_3$ (CTO), characterized by the ABX$_{3}$ configuration. Its unit cell is a cube of space group $Pm\bar{3}m~(221)$, with cations A and B located at the vertices and body centers, respectively, and anion X at the face centers, as illustrated in Fig.~\ref{struc}(a). The cation A acts as a fixed shell, making its chemical and physical properties relatively less significant. However, the displacement of the B and X atoms, breaking the centrosymmetry, transforms the cube into a tetrahedral structure inducing spontaneous polarization. This displacement, which governs the FE properties of perovskites, can be controlled by modifying the electronic configuration of the cation B through geometric changes in the BX$_6$ octahedron.\cite{Tilley2016} Typically, A represents an alkali or alkaline earth metal, while B represents a transition metal element.\cite{wei2020energy} In this study, we considered only divalent (Group 2: Ca, Sr, Ba, Ra) and tetravalent (Group 4: Ti, Zr, Hf) elements for cations A and B, respectively, with oxygen as anion X.

To investigate the effect of FE materials on spin switching in FM materials, we considered iron (Fe), a prime example of a magnetic material, as an adsorbate on ABO$_3$. To gain a fundamental understanding of Fe adsorption on ABO$_3$, we first investigated its interaction with barium titanate, BaTiO$_3$ (BTO), a well-established FE material. Following the experimental observations revealing the interfacial structure between iron and BTO~\cite{radaelli2014electric} and iron adsorption at the atomic scale,~\cite{wagner2016, Nanoscale2020, Fan2021}, we constructed a model structure of Fe-adsorbed BTO as follows. We created a BTO surface with the TiO$_2$ layer and then added iron atoms to the surface one by one until a full monolayer coverage ($\Theta=1$) was achieved. Fig.~\ref{struc}(b) shows two configurations of Fe adsorption on BTO at coverage of 50\% ($\Theta=1/2$) and 100\% ($\Theta=1$). Note that we only considered the positive BTO polarization, under which Fe atoms are initially more readily absorbed on the B top site than under its negative polarization.

To quantitatively investigate the preferred adsorption sites for Fe atoms, we calculated the Fe atom adsorption energy $E_\mathrm{ads}$, defined as
\begin{equation}
E_\mathrm{ads} = E_\mathrm{tot}^\mathrm{Fe/BTO}-E_\mathrm{tot}^\mathrm{BTO}-E_\mathrm{tot}^\mathrm{Fe},
\end{equation}
where $E_\mathrm{tot}^\mathrm{Sys}$ is the total energy of a system Sys that is Fe/BTO, BTO, or Fe, denoting the system of the BTO substrate with a Fe adsorbate, the BTO (001) substrate alone, or the Fe adsorbate alone, respectively. Fig.~\ref{struc}(c) shows the calculated $E_\mathrm{ads}$ as a function of the coverage, where the red and blue curves represent two cases of Fe adsorption on the Ba and O top sites, respectively. For $\Theta\le1/2$, the iron atoms prefer the Ba top sites, and the O atoms at the interface have been placed between the adsorbed Fe atoms. At $\Theta=1/2$, the adsorbed Fe atoms occupy all available Ba top sites and form a complete two-dimensional FeO$_2$ monolayer. At higher coverages, however, it seemed that the next Fe atom would have no choice but to sit on one of the Ti top sites, unless the Fe atoms already adsorbed on the Ba top site and the O atoms in the FeO$_2$ layer rearranged their positions. However, at $\Theta=5/8$, shown in the green shared region in Fig.~\ref{struc}(c), the newly added Fe atoms do not adsorb on the Ti top site, but rather tend to push the O atoms down and displace the Fe atoms already adsorbed on the Ba top site toward the O top site, forming an iron cluster, as in the configuration at 62.5\% shown in Fig.~\ref{struc}(c), which significantly lowers the system energy. At higher coverages, the adsorbed Fe atoms form a tightly packed square lattice on the O top sites. This results in a distinct interface between the Fe layer and the BTO substrate, separated by the TiO$_2$ layer. Note that further Fe additions lead to forming the bulk-like Fe layers with oxidation occurring at the interface.\cite{Valencia2011,radaelli2014electric} To understand the effect of FE substrates on the iron magnetic behaviors, we focused on the Fe adsorbate with $\Theta=1$ on various ABO$_3$, in which we could safely exclude the complex effects of oxidation at the interface. Fortunately enough, moreover, various ABO$_3$ compounds have a very similar lattice constant to that of BTO within less than $\pm5$\% and share similar chemical properties for Fe adsorption behaviors. It should be noted that the Ca-based perovskites experience octahedral rotation,\cite{Kim2020,marcondes2021stability} resulting in different structural geometry to other alkaline-earth element-based ones. Nonetheless, we maintained the same structure even for the Ca-based oxides in our study. Introducing octahedral rotation would complicate the systematic understanding of interfacial effects. Additionally, it has been shown that the octahedra in ABO$_3$ compounds can remain unrotated by adjusting various conditions such as doping, strain, and defects.\cite{Biegalski2015,Herklotz2016,Jia2022} Therefore, maintaining an unrotated octahedral structure allows for a clearer analysis of the interfacial phenomena under investigation.

\begin{figure}[t]
\centering
\includegraphics[width=1.0\columnwidth]{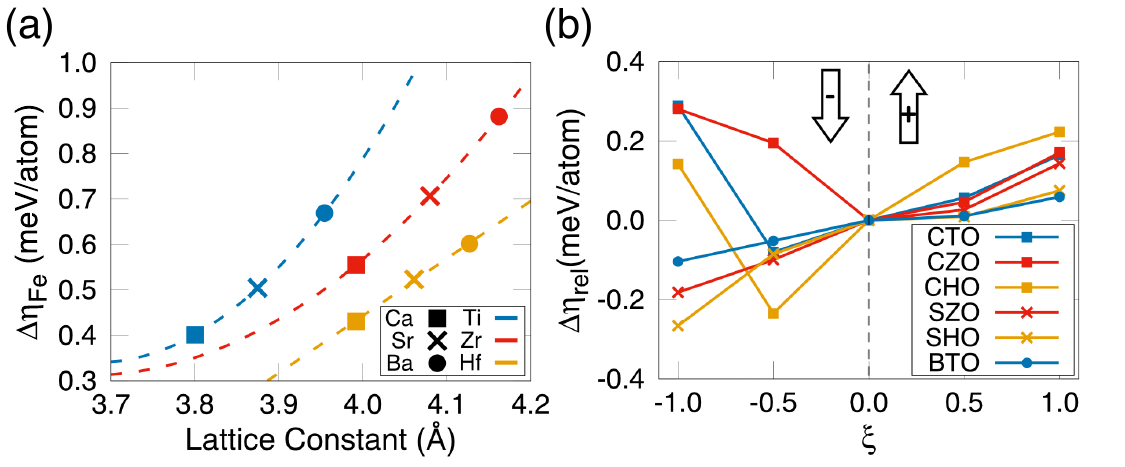}
\caption{(a) The relative MAE $\Delta\eta_\mathrm{Fe}$, defined in Eq.~\ref{eta_Fe}, of iron adsorbates on different ABO$_3$ substrates (A = Ca, Sr, or Ba; B = Ti, Zr, or Hf) at full coverage. The divalent elements Ca, Sr, and Ba corresponding to the cation A are represented by solid squares, crosses, and solid circles, respectively. The tetravalent elements Ti, Zr, and Hf for the cation B are represented by blue, red, and orange colors, respectively. Dashed lines indicate quadratic fits for each B element, highlighting the classification by tetravalent elements. (b) The relative MAE $\Delta\eta_\mathrm{rel}$, defined in Eq.~\ref{eta_rel}, as a function of $\xi=P/P_\mathrm{s}$ for six ABO$_3$ substrates labeled with the same symbol and color as in (a). The vertical dashed line marks $\xi=0$. Three ABO$_3$ substrates with zero spontaneous polarization are excluded. The linear behavior in SZO, SHO, and BTO contrasts with the unexpected upturns in CTO, CZO, and CHO as $\xi$ decreases.
\label{MAE}}
\end{figure}
We utilized the slab model, as shown in Fig.~\ref{struc}(b), to examine the interplay between the MAE and the spontaneous polarization of various ABO$_3$ materials. To achieve an equilibrium interfacial structure, we fully relaxed the top three layers of the ABO$_3$ substrate and the adsorbed Fe atoms within the same crystal symmetry. The bottom four layers were fixed to preserve the bulk structure to clearly capture the effect of spontaneous polarization on the MAE of the adsorbed Fe. The polarization direction depends on the relative displacement of BO$_6$ octahedra within the ABO$_3$ unit cell. When the octahedra is displaced towards the adsorbed Fe layer, the polarization becomes negative. Conversely, displacement in the opposite direction leads to positive polarization.
Having established the interface between the adsorbed Fe and ABO$_3$ substrate, we first studied the MAE of the Fe adsorbate on the unpolarized ABO$_3$ structures with the centrosymmetric structure shown in Fig.~\ref{struc}(a). The MAE $\eta$ is the energy required to change the direction of magnetization from the out-of-plane direction to the in-plane direction in the magnetic thin film material, defined as
\begin{equation}
\eta = E_\mathrm{tot}(S_\rightarrow)-E_\mathrm{tot}(S_\uparrow),
\label{eq1}
\end{equation}
where $E_\mathrm{tot}(S_\rightarrow)$ and $E_\mathrm{tot}(S_\uparrow)$ are the total energies for the systems with the in-plane ($S_\rightarrow$) and out-of-plane ($S_\uparrow$) spin configurations, respectively. A positive (negative) value of the MAE means that the spin prefers the out-of-plane (in-plane) direction. The MAE arises from the SOC effect, where the orbitals of atoms affected by the crystal field interact with spins.\cite{daalderop1990first} So, it is also called magneto-crystalline anisotropy energy (MCA). To explore the effects of FE substrates alone on the MAE, excluding the strain effect, we evaluated $\Delta\eta_\mathrm{Fe}$ defined as
\begin{equation}
\Delta\eta_\mathrm{Fe}=\eta_\mathrm{Fe/ABO_3}-\eta_\mathrm{Fe^{free}_s},
\label{eta_Fe}
\end{equation}
where $\eta_\mathrm{X}$ is the MAE of a system X, which is either Fe/ABO$_3$ or Fe$\mathrm{^{free}_s}$, representing a slab system of adsorbed Fe layer on an ABO$_3$ substrate or the same Fe layer without the ABO$_3$ substrate, respectively. The range of $\eta_\mathrm{Fe/ABO_3}$ is from 0.57 to 0.86~meV/atom, which indicates that all Fe/ABO$_3$ systems prefer the out-of-plane spin configuration. The estimated $\Delta\eta_\mathrm{Fe}$ on various ABO$_3$ substrates is 0.40 to 0.88~meV/atom, as shown in Fig.~\ref{MAE}(a).
It is shown that for a given B element of ABO$_3$ substrate, the heavier the A element, the higher the MAE. The MAE data were well fitted to a quadratic function of the lattice constant of ABO$_3$ for each B element, as shown in Fig.~\ref{MAE}(a).

To explore the inherent effect of the spontaneous polarization of ABO$_3$ substrate alone on the MAE of the adsorbed Fe, we also evaluated the relative MAE $\Delta\eta_\mathrm{rel}$ using 
\begin{equation}
\Delta{\eta}_\mathrm{rel}={\eta(\xi)-\eta(0)},
\label{eta_rel}
\end{equation}
where $\xi$ represents the normalized atomic displacement corresponding to the relative polarization ($P$) of each ABO$_3$ with respect to its spontaneous polarization ($P_\mathrm{s}$), or $\xi=P/P_\mathrm{s}$. For example, $\eta(0)$ denotes the MAE of the Fe placed on the corresponding centrosymmetric ($\xi=0$) ABO$_3$ substrate with zero polarization. Fig.~\ref{MAE}(b) presents our evaluated $\Delta\eta_\mathrm{rel}$ as a function of $\xi$. It is noted that negative values in Fig.~\ref{MAE}(b) do not indicate that the spins on the Fe adsorbate prefer the in-plane direction but rather demonstrate the change in MAE with respect to their unpolarized cases, in which the Fe adsorbate still favors the out-of-plane spins. In this figure, we did not include three ABO$_3$ materials, STO, BZO, and BHO, since they are not FE, indicating zero spontaneous polarization. Our results demonstrate that the MAE of the adsorbed Fe is linearly dependent on $\xi$ on SZO, SHO, and BTO, as naturally expected that the MAE would be affected by the electric field induced by the spontaneous polarization of the substrate. However, when the Fe adsorbate is placed on the other three substrates containing Ca element for the A-site, we observed unusual upturns in $\Delta\eta_\mathrm{rel}$ as $\xi$ decreases.

\begin{figure*}[t]
\centering
\includegraphics[width=0.9\textwidth]{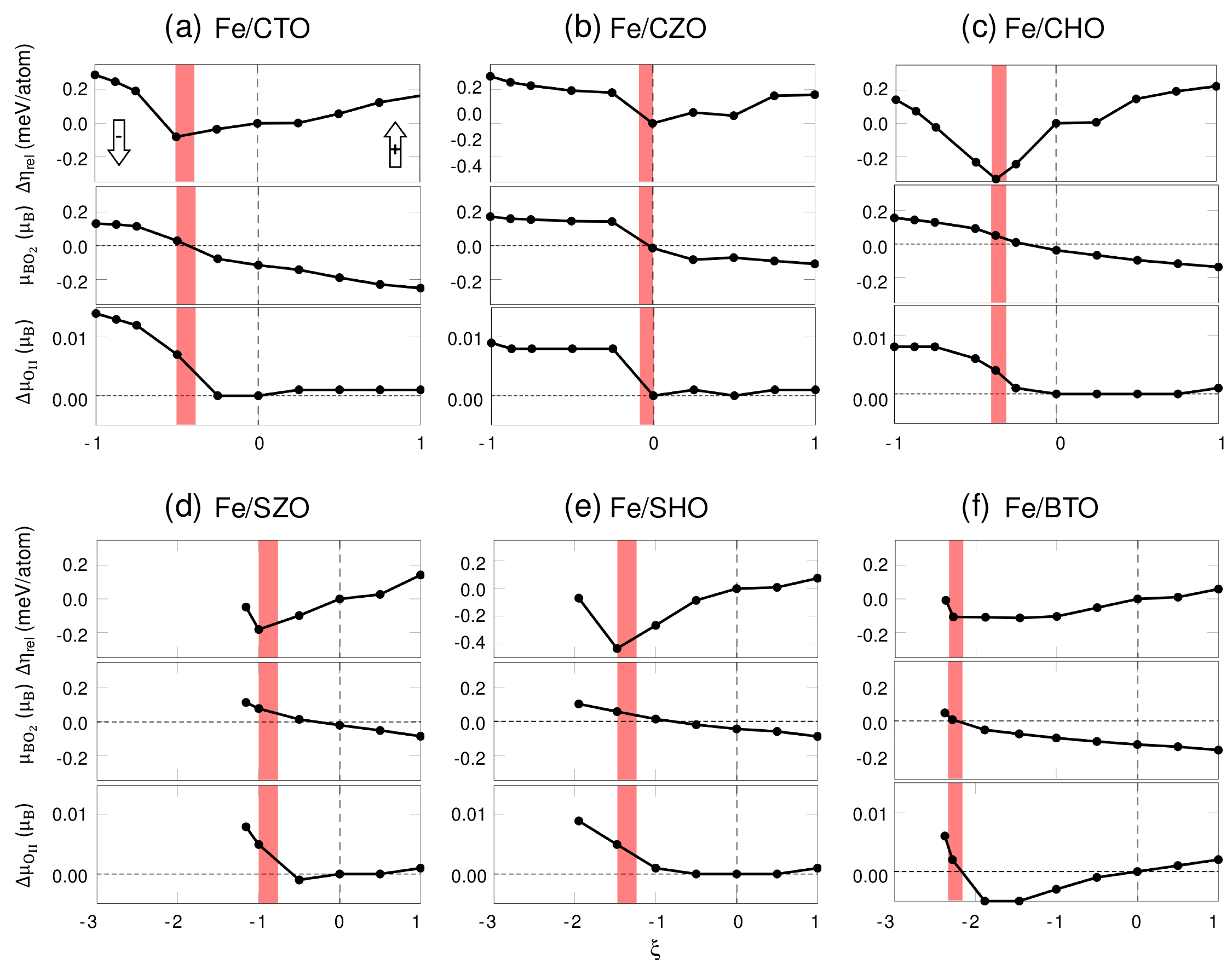}
\caption{The relative MAE $\Delta\eta_\mathrm{rel}$, induced magnetic moments $\mu_\mathrm{BO_{2}}$ in the BO$_2$ layer, and relative magnetic moments $\Delta\mu_\mathrm{O_{II}}$ induced at O$_\mathrm{II}$ atoms for (a) Fe/CTO, (b) Fe/CZO, (c) Fe/CHO, (d) Fe/SZO, (e) Fe/SHO, and (f) Fe/BTO as a function of $\xi$. These relative quantities were evaluated with respect to their corresponding values at $\xi=0$. A sharp sign change in the slope of $\Delta\eta_\mathrm{rel}$ in the region highlighted in red correlates strongly with changes in $\mu_\mathrm{BO_{2}}$ and $\Delta\mu_\mathrm{O_{II}}$. Notice the $\xi$ range in (d--f) differs from that in (a--c), intentionally showing that the sign change in the slope occurs beyond the spontaneous polarization in the three cases with heavier A elements shown in (d--f).
\label{MMO2}}
\end{figure*}

We now turn our attention to the abrupt upturns in $\Delta\eta_\mathrm{rel}$ observed in three cases (Fe/CTO, Fe/CZO, and Fe/CHO) as shown in Fig.~\ref{MAE}(b). Notably, these upturns occur within the range of spontaneous polarization ($|\xi| < 1$). In contrast, the other three cases (Fe/SZO, Fe/SHO, and Fe/BTO) do not exhibit such upturns within the same polarization range. This disparity suggests the presence of factors beyond the electric field effects induced by spontaneous polarization. Specifically, our analysis reveals that the MAE upturns are predominantly influenced by interface effects, such as induced interfacial magnetic dipole moments, as described in the following. To illustrate this interface-driven effect, we analyzed how the electronic structure evolves with polarization, as detailed in Supplementary Note S2. Supplementary Fig.S3 represents the projected density of states (PDOS) of Fe $3d$ orbitals under varying polarization states. In CTO, polarization reversal leads to marked shifts and reshaping of Fe $3d$ peaks, indicating a strong change in orbital character and hybridization. In contrast, BTO exhibits only minimal changes in Fe $3d$ states. This distinction aligns with the sharp MAE jump observed in CTO and its absence in BTO, confirming that polarization-induced modification of Fe–O–metal bonding is the key factor controlling magnetic anisotropy. While LDA may underestimate absolute band alignment, it reliably captures the polarization-driven modulation of Fe orbital anisotropy that governs  the observed trends.

Fig.~\ref{MMO2} presents the relative MAE ($\Delta\eta_\mathrm{rel}$), the induced magnetic moments of the interface BO$_2$ layer ($\mu_\mathrm{BO_2}$), and the relative magnetic moment induced in the oxygen atom O$\mathrm{II}$ ($\Delta\mu_\mathrm{O_{II}}$), located just below the topmost B atom in ABO$3$ as depicted in Fig.~\ref{struc}(b), as functions of $\xi$ for all six cases. Abrupt upturns in $\Delta\eta\mathrm{rel}$ are observed in the region $\xi < 0$ for all cases. Notably, in the first three cases (Fe/CTO, Fe/CZO, and Fe/CHO), these upturns occur within $-1 < \xi < 0$, whereas in the other three cases (Fe/SZO, Fe/SHO, and Fe/BTO), the upturns are observed only when $\xi < -1$, beyond the spontaneous polarization range. We observed that $\mu_\mathrm{BO_2}$ changes its sign near $\xi_c$, the critical $\xi$ value where the upturns occur. Furthermore, $\Delta\mu_\mathrm{O_{II}}$ exhibits a sharp change near $\xi_c$, transitioning from positive for $\xi<\xi_c$ to near zero for $\xi>\xi_c$, mirroring the behavior of MAE. This emphasizes that, in practical scenarios, MAE upturns are apparent only in the first three cases within their spontaneous polarization limits. For the other three cases, external electric fields are necessary to extend polarization beyond the $\xi=-1$ threshold to observe similar effects.

Additionally, we found that as the A-site element becomes heavier, the critical value $\xi_c$ shifts further into the negative range. This trend is attributed to the increased stability of the octahedron in ABO$_3$ perovskites with heavier A-site elements, leading to reduced octahedral deformation. A critical degree of octahedral deformation appears to be essential for significantly enhancing the interfacial magnetic moments of both B and O$_\mathrm{II}$ atoms, ultimately driving the observed MAE upturns. Our analysis underscores the essential role of changes in the magnetic moments of interface atoms induced by spontaneous polarization. These findings highlight the complex interplay of structural and magnetic properties in determining the MAE of adsorbed Fe in ABO$_3$ perovskites, providing valuable insights into their multi-dimensional effects.

Building on these findings, our analysis identifies two distinct factors that contribute to the MAE of adsorbed Fe in ABO$_3$ systems influenced by spontaneous polarization. The first factor is the ``interface effect'' arising from atomic displacements induced by ferroelectricity. To elucidate the atomic-level mechanism of the MAE jump, we find that polarization switching induces vertical displacements of interfacial oxygen atoms relative to the Fe layer. This structural shift modifies the Fe–B and B–O$_\mathrm{II}$ bond lengths, thereby altering the Fe–B orbital hybridization. Since orbital hybridization determines the orbital moment anisotropy of Fe atoms, these polarization-driven adjustments at the interface ultimately lead to the observed discrete jump in MAE. As shown in Supplementary Fig.S2, when $\xi$ assumes a negative value, the separation between the Fe and B atoms increases. The magnetic moment of the B atom in the BO$_2$ layer, which opposes that of the adsorbed Fe, decreases significantly in magnitude, by approximately $0.2$ to $0.4~\mu_B$ in the negative $\xi$ regime. In contrast, the magnetic moment of the O atoms in the same plane, which aligns with the Fe moment, exhibits a comparatively smaller increase, typically $0.02$ to $0.04~\mu_B$. This difference highlights the dominant influence of the B atoms in converting the net magnetic moment of the BO$_2$ layer to a positive value, which amplifies the MAE by reinforcing the out-of-plane anisotropy of the Fe moment. Furthermore, the displacement of atoms in the negative $\xi$ regime brings the O$_\mathrm{II}$ atom closer to the B atom, as shown in Fig.~\ref{struc}(b), inducing a magnetic moment in O$_\mathrm {II}$ through the proximity effect. The combined movement of the B atom and the reduced magnetic moment at the interface leads to the induction of a magnetic moment in O$_\mathrm{II}$, which correlates with the observed abrupt upturns in the MAE of the adsorbed Fe. The second factor is associated with the electric field directly induced by spontaneous polarization. However, this effect is relatively minor, as detailed in Supplementary Note S1.

\begin{figure}[t]
\includegraphics[width=1.0\columnwidth]{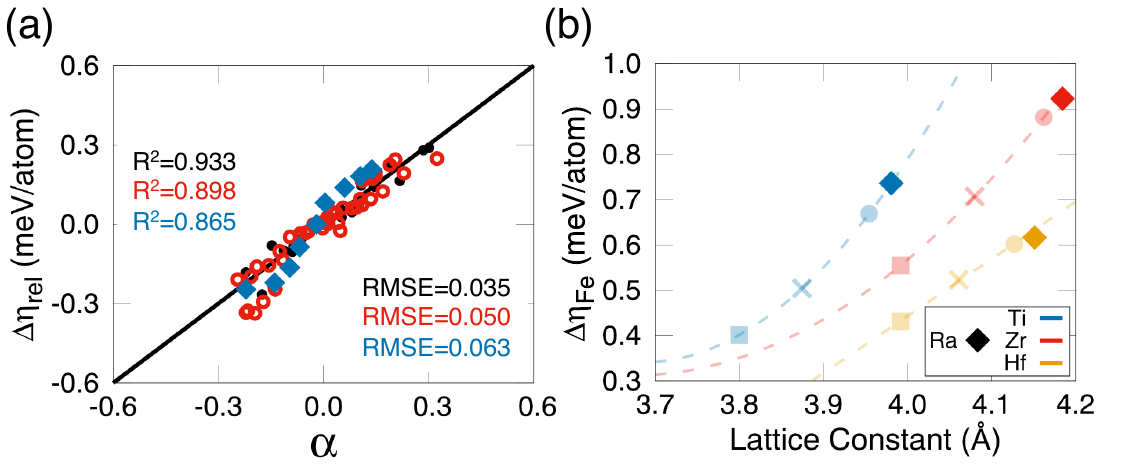}
\caption{(a) Relationship between the relative MAE $\Delta\eta_\mathrm{rel}$ and the identified descriptor $\alpha$ given in Eq.~\ref{alpha}. Solid black circles denote the training set ($R^2= 0.933$), empty red circles the validation set ($R^2=0.898$), and blue diamonds the test set ($R^2=0.865$). (b) The relative MAE $\Delta\eta_\mathrm{Fe}$ of Fe-adsorbed RaBO$_3$ substrates (B$=$Ti, Zr, Hf) obtained from the identified descriptor $\alpha$ given in Eq.~\ref{alpha}, which were added to Fig.~\ref{MAE}(a), showing that our machine learning descriptor can also be applied to the heavier A = Ra element in ABO$_3$.
\label{ML}}
\end{figure}

We employed the SISSO scheme for machine learning to identify key parameters for accurately representing $\Delta\eta_\mathrm{rel}$. Training the model on a 30-set dataset excluding unpolarized ABO$_3$, we evaluated features such as the ABO$_3$ lattice constant, polarization value, and the magnetic moments of individual atoms in the Fe adsorbate-ABO$_3$ system under spontaneous polarization. 
These primary features were selected based on their direct physical relevance to MAE. The magnetic moment of Fe atoms directly influences MAE. The ABO$_3$ lattice constant indirectly affects local strain and site symmetry at the Fe adsorption site and thus alters the electronic structure and magnetic properties. The polarization value changes the interfacial electric field and atomic displacements, as it is widely considered a key factor that could influence MAE. Based on this physically informed selection, the SISSO algorithm systematically constructed composite descriptors and identified the optimal combination for the best correlation with the evaluated MAE values. This analysis yielded the following descriptor,
\begin{equation}
\alpha=\left(\mu_\mathrm{B}+\mu_\mathrm{Fe}\right)\left(\lambda_1\mu_\mathrm{O_{II}}\mu_\mathrm{Fe}+\lambda_{2}a^6\right)+\gamma,
\label{alpha}
\end{equation}
where, $\alpha$ denotes the descriptor, $\lambda_{1,2}$ are coefficients valued at $1.814\times10^3$ and $4.312\times10^{-4}$, respectively, and $\gamma$ is a constant valued at $1.951\times10^{-2}$. The parameters $a$, $\mu_\mathrm{B}$, $\mu_\mathrm{Fe}$, and $\mu_\mathrm{O_{II}}$ represent the lattice constant and the magnetic moments of the B site, Fe atom, and O$\mathrm{II}$ atom, respectively. Notably, this two-dimensional descriptor, despite its simplicity, effectively captures $\Delta\eta_\mathrm{rel}$ by incorporating the magnetic moments of Fe, B, and O$_\mathrm{II}$ atoms alongside the lattice constant.

Fig.~\ref{ML}(a) illustrates the high predictive accuracy of $\alpha$ for the training set of Fe adsorbates on ABO$_3$ substrates (solid black circles), achieving $R^2=0.933$ and $\mathrm{RMSE}=0.035$, where RMSE denotes root-mean-squared error. Furthermore, the descriptor demonstrates strong predictive power for untrained data (empty red circles) with $R^2=0.898$ and $\mathrm{RMSE}=0.050$. Even when applied to an extrapolated dataset of RaTiO$_3$, which is absent from the training set, it maintains robust accuracy ($R^2=0.865$, $\mathrm{RMSE}=0.063$), as indicated by the blue rhombi. The predicted MAE for RaBO$_3$ (B = Ti, Zr, Hf) aligns closely with the trends observed in the training dataset, as depicted in Fig.~\ref{ML}(b), which is the same as Fig.~\ref{MAE}(a) with additional data points for RaBO$_3$. These results underscore the utility of our machine learning approach for predicting MAE values in diverse ABO$_3$ compounds with Fe adsorbates. By capturing the complex interplay of interfacial magnetic moments and lattice effects, our descriptor provides a powerful tool for guiding the design of advanced ferroelectric and spintronic materials. Furthermore, the predicted discrete jump in MAE driven by interfacial structural changes is expected to be experimentally detectable via techniques such as X-ray magnetic circular dichroism (XMCD)\cite{radaelli2014electric} or ferromagnetic resonance (FMR)\cite{tortarolo2025charge}, providing a viable pathway for experimental validation of our theoretical predictions.
 
\section{Conclusions}
In summary, this study addresses the challenge of controlling the magnetic anisotropy energy (MAE) in ferromagnetic (FM) materials to enable low-power spin-based devices. By exploring the influence of ferroelectric (FE) spontaneous polarization, we investigated ABO$_3$ materials as substrates and their effects on Fe adsorbates. Our results revealed significant variations in MAE across ABO$_3$ compounds with different A and B elements. A key finding was the unexpected jump in MAE, which was primarily driven by interface effects rather than the electric field effects associated with spontaneous polarization. This discovery underscores the critical role of the interplay between the Fe adsorbate and the ABO$_3$ substrate at the atomic level. Leveraging machine learning, we developed a descriptor that accurately captures the physical factors affecting MAE. By incorporating the magnetic moments of interface atoms and the lattice constant, our model achieved high predictive accuracy, with $R^2=0.933$ and $\mathrm{RMSE}=0.035$ on the training set. Notably, the model demonstrated strong performance when extrapolated to unseen data, including RaTiO$_3$, achieving $R^2=0.865$ and $\mathrm{RMSE}=0.063$. These results validate the robustness and generalizability of our approach, offering a predictive tool for tailoring MAE in a wide range of ABO$_3$ compounds. Our study also highlights how the selection of A and B elements in perovskite structures significantly influences MAE. Heavier A elements, such as Sr and Ba, were found to stabilize the octahedral structure of ABO$_3$, reducing deformation and modulating the MAE more effectively. This insight provides a pathway for engineering materials that exhibit optimal MAE characteristics, enabling faster and more energy-efficient switching in spintronic devices. For example, incorporating ABO$_3$ compounds with A~=~Ca into device architectures could reduce magnetization switching times by exploiting the MAE jump, without requiring changes to the external electric field direction.

Overall, these findings advance our understanding of the mechanisms underlying MAE modulation in FM materials on FE materials and establish a framework for designing high-performance spintronic devices. By bridging first-principles calculations and machine learning, this work contributes to the broader field of materials science, offering strategies for the development of next-generation electronic and spintronic technologies.

\section{Acknowledgements}
This research was supported by the Korean government (MSIT) through the National Research Foundation of Korea (NRF-2022R1A2C1005505, NRF-2022M3F3A2A01073562, and NRF-RS-2024-00416976) and Institute for Information $\&$ Communications Technology Planning $\&$ Evaluation (IITP) (2021-0-01580). S.-H. K. is supported by the US Department of Energy (DOE), Office of Science, National Quantum Information Science Research Centers, Quantum Science Center. Our computational work was done using the resources of the KISTI Supercomputing Center (KSC-2023-CRE-0053 and KSC-2024-CRE-0211) and the resources of the Oak Ridge Leadership Computing Facility, US DOE Office of Science User Facilities.

This manuscript has been coauthored by UT-Battelle, LLC, under Contract No. DE-AC0500OR22725 with the U.S. Department of Energy. The United States Government retains and the publisher, by accepting the article for publication, acknowledges that the United States Government retains a non-exclusive, paid-up, irrevocable, world-wide license to publish or reproduce the published form of this manuscript or allow others to do so, for the United States Government purposes. The Department of Energy will provide public access to these results of federally sponsored research in accordance with the DOE Public Access Plan (http://energy.gov/downloads/doe-public-access-plan).



\bibliographystyle{rsc}
\bibliography{rsc} 

\end{document}